\documentclass[12pt]{article}
\usepackage{amsmath,amssymb}
\usepackage{bm}
\usepackage{mathrsfs}
\usepackage{fourier}
\usepackage{multirow}
\allowdisplaybreaks[4]
\newcommand{{\Slashp}}{p\!\!\!\!\!\big/}
\newcommand{{\Slashq}}{q\!\!\!\!\!\big/}
\usepackage[usenames]{color}

\begin{document}

\title{Family number, Wilson line phases\\
and hidden supersymmetry}

\author{
Yuhei \textsc{Goto}\footnote{E-mail: 14st302a@shinshu-u.ac.jp},~~~
Yoshiharu \textsc{Kawamura}\footnote{E-mail: haru@azusa.shinshu-u.ac.jp}\\
{\it Department of Physics, Shinshu University, }\\
{\it Matsumoto 390-8621, Japan}\\
and \\
Takashi \textsc{Miura}\footnote{E-mail: takashi.miura@people.kobe-u.ac.jp}\\
{\it Department of Physics, Kobe University, }\\
{\it Kobe 657-8501, Japan}\\
}

\date{
May 27, 2014}

\maketitle
\begin{abstract}
We study the relationship between the family number of chiral fermions
and the Wilson line phases, based on the orbifold family unification.
We find that flavor numbers are independent 
of the Wilson line phases relating extra-dimensional components of gauge bosons,
as far as the standard model gauge symmetry is respected.
This feature originates from a hidden quantum-mechanical supersymmetry.
\end{abstract}

\section{Introduction}
\label{Introduction}

The origin of the chiral fermions and the family replication has been a big mystery.
On a higher-dimensional space-time including an orbifold as an extra space,
the family unification based on a large gauge group or the structure of extra dimensions
can provide a possible 
solution~\cite{BB&K,W&T,GMN1,GMN2,GLMN,GLMS,KK&O,M&N,K&M,FKMN&S,GK&M,AFKMN&S}.

Four-dimensional (4D) chiral fermions come from a higher-dimensional fermion,
after the elimination of mirror particles by orbifolding upon compactification.
The symmetry breaking mechanism on the orbifold has been originally used
in superstring theory~\cite{DHV&W1,DHV&W2}.
The family replication emerges from a few multiplets of a large gauge group
including the family group as a subgroup.
Hence, it is interesting to explore a nature of the family number 
based on the orbifold family unification, 
in the expectation that it offers a hint on the origin of three families
in the standard model (SM).

In this paper, we study the relationship between the family number of chiral fermions
and the Wilson line phases, based on the orbifold family unification.
We find that flavor numbers are independent of the Wilson line phases
relating extra-dimensional components of gauge bosons,
as far as the SM gauge symmetry is respected.
This feature originates from a hidden quantum-mechanical supersymmetry (SUSY).

The contents of this paper are as follows.
In Sec. II, we review a feature of the family number in the orbifold family unification,
and present a conjecture on flavor numbers.
In Sec. III, we give an example to support the conjecture 
and show that it is understood from the viewpoint of the hidden SUSY.
Section IV is devoted to conclusions.
In Appendices A and B, we present several formulas concerning the combination ${}_n C_{l}$,
derived from the feature that flavor numbers are independent of the Wilson line phases.

\section{Family number in orbifold family unification}
\label{Family number in orbifold family unification}

\subsection{A feature}
\label{A feature}

In our previous work~\cite{GK&M}, we have studied the orbifold family unification 
in $SU(N)$ gauge theory on 6D space-time, $M^4 \times T^2/Z_M$ $(M=2,3,4,6)$.
Here, $M^4$ is the 4D Minkowski space-time and $T^2/Z_M$ is the 2D orbifold.
We have derived enormous numbers of models with three families 
of $SU(5)$ multiplets and the SM multiplets 
from a pair of 6D Weyl fermions with different chiralities, 
using the orbifold breaking mechanism,
after the breakdown of gauge symmetry such that
$SU(N) \rightarrow SU(5) \times SU(p_2) \times \cdots \times SU(p_n) \times U(1)^{n-1-m}$ 
and $SU(N) \rightarrow SU(3) \times SU(2) \times 
SU(p_3) \times \cdots \times SU(p_n) \times U(1)^{n-1-m}$, respectively.
Here and hereafter, $m$ is the number of zero in $\{p_i\}$ 
and ``$SU(1)$'' unconventionally stands for $U(1)$, $SU(0)$ means nothing.

Through the analysis, we have found the feature that
 {\it each flavor number obtained from a 6D Weyl fermion with $[N,k]$
is invariant under the change $\{p_i\}$ into $\{p'_i\}$
among the equivalent boundary conditions (BCs).}
Here, $[N, k]$ is the rank $k$ totally antisymmetric tensor representation
whose dimension is ${}_N C_k$.

Let us present several illustrations.

On $T^2/Z_2$, the numbers of 4D left-handed Weyl fermions
with the representations $\overline{\bm{5}}$ and $\bm{10}$
obtained from the 
breaking pattern $SU(N) \to  SU(5) \times SU(p_2) \times \cdots \times SU(p_8) \times U(1)^{7-m}$
are same as those from
$SU(N) \to  SU(5) \times SU(p'_2) \times \cdots \times SU(p'_8) \times U(1)^{7-m}$,
if the following relations are satisfied,
\begin{align}
&p'_2 - p_2 = p'_7 - p_7 = p_3 - p'_3 = p_6 - p'_6~, \notag \\
&p'_4 = p_4~,~~ p'_5 = p_5~,~~ p'_8 = p_8~,
\label{equ-r1}
\end{align}
or 
\begin{align}
&p'_2 - p_2 = p'_7 - p_7 = p_4 - p'_4 = p_5 - p'_5~,\notag \\
&p'_3 = p_3~,~~ p'_6 = p_6~,~~ p'_8 = p_8~,
\label{equ-r2}
\end{align}
or
\begin{align}
&p'_3 - p_3 = p'_6 - p_6 = p_4 - p'_4 = p_5 - p'_5~,\notag \\
&p'_2 = p_2~,~~ p'_7 = p_7~,~~p'_8 = p_8~.
\label{equ-r3}
\end{align}

In the same way, the flavor numbers of the SM fermions 
obtained from
$SU(N) \to SU(3) \times SU(2) \times SU(p_3) \times \cdots \times SU(p_8) \times U(1)^{7-m}$
are same as those from
$SU(N) \to SU(3) \times SU(2) \times SU(p'_3) \times \cdots \times SU(p'_8) \times U(1)^{7-m}$,
if the following relations are satisfied,
\begin{align}
p'_3 - p_3 = p'_6 - p_6 = p_4 - p'_4 = p_5 - p'_5~,~~ p'_7 = p_7~,~~
p'_8 = p_8~.
\label{equ-r1-SM}
\end{align}

On $T^2/Z_3$, the numbers of 4D left-handed Weyl fermions
with $\overline{\bm{5}}$ and $\bm{10}$ obtained from 
$SU(N) \to  SU(5) \times SU(p_2) \times \cdots \times SU(p_9) \times U(1)^{8-m}$
are same as those from
$SU(N) \to  SU(5) \times SU(p'_2) \times \cdots \times SU(p'_9) \times U(1)^{8-m}$,
if the following relations are satisfied,
\begin{eqnarray}
p'_2 - p_2 = p'_6 - p_6 = p'_7 - p_7 = p_3 - p'_3 = p_4 - p'_4 = p_8 - p'_8~,~~
p'_5 = p_5~,~~p'_9 = p_9~.
\label{equ-r1-Z3}
\end{eqnarray}

\subsection{A conjecture}
\label{A conjecture}

In the above cases with the relations (\ref{equ-r1}) -- (\ref{equ-r1-Z3}),
the BCs relating $\{p_i\}$ 
are connected with those relating $\{p'_i\}$ 
by singular gauge transformations,
and they are regarded as equivalent in the presence of the Wilson line phases
relating extra-dimensional components of gauge bosons.
This equivalence originates from the dynamical rearrangement
in the Hosotani mechanism~\cite{Hosotani1,Hosotani2,HHH&K,HH&K}.

For cases on $T^2/Z_2$, the equivalence of BCs is shown by the following relations 
among the diagonal representatives 
for $2 \times 2$ submatrices of $(P_0, P_1, P_2)$~\cite{K&M2},
\begin{align}
(\tau_3, \tau_3, \tau_3) \sim (\tau_3, \tau_3, -\tau_3) \sim (\tau_3, -\tau_3, \tau_3)
\sim (\tau_3, -\tau_3, -\tau_3)~,
\label{equ-Z2}
\end{align}
where $P_0$, $P_1$ and $P_2$ are representation matrices for the $Z_2$ reflections, and
$\tau_3$ is the third component of the Pauli matrices.
For case on $T^2/Z_3$, it is shown by the following relations among the diagonal representatives 
for $3 \times 3$ submatrices of $(\Theta_0, \Theta_1)$~\cite{K&M2},
\begin{eqnarray}
(X, X) \sim (X, \overline{\omega} X) \sim (X, \omega X)~,
\label{equ-Z3}
\end{eqnarray}
where  $\Theta_0$ and $\Theta_1$ are representation matrices for the $Z_3$ rotations,
and $X = \mathrm{diag}(1, \omega, \overline{\omega})$ with
$\omega = e^{2\pi i/3}$ and $\overline{\omega} = e^{4\pi i/3}$.

In \cite{GK&M}, we assume that the BCs are chosen as physical ones, i.e., the system with
the physical vacuum is realized with the vanishing Wilson line phases
after a suitable gauge transformation is performed.
Then, the feature is expressed by 
\begin{eqnarray}
  \left. N_{\bm{r}} \right|_{(\{ p_i \},a_k=0)} 
 = \left. N_{\bm{r}} \right|_{(\{ p'_i \},a_k=0)}~,
\label{Nr1}
\end{eqnarray}
where $N_{\bm{r}}$ is a net chiral fermion number (flavor number) for 4D fermions
with the representation $\bm{r}$ of the gauge group, 
unbroken even in the presence of the Wilson line phases $(2\pi a_k)$,
and it is defined by
\begin{eqnarray}
N_{\bm{r}} \equiv n^0_{{\rm L}\bm{r}} - n^0_{{\rm R}\bm{r}}
- n^0_{{\rm L}\overline{\bm{r}}} + n^0_{{\rm R}\overline{\bm{r}}}~.
\label{Nr-def}
\end{eqnarray} 
Here, $n^0_{{\rm L}\bm{r}}$, $n^0_{{\rm R}\bm{r}}$,
$n^0_{{\rm L}\overline{\bm{r}}}$ and $n^0_{{\rm R}\overline{\bm{r}}}$
are the numbers of 4D left-handed massless fermions with $\bm{r}$,
4D right-handed one with $\bm{r}$, 4D left-handed one with the complex conjugate representation
$\overline{\bm{r}}$ and 4D right-handed one with $\overline{\bm{r}}$, respectively.
Note that 4D right-handed fermion with $\overline{\bm{r}}$
and 4D left-handed one with $\bm{r}$
are transformed into each other under the charge conjugation.

On the other hand, the equivalence due to the dynamical rearrangement is expressed by 
\begin{eqnarray}
  \left. N_{\bm{r}} \right|_{(\{ p_i \},a_k \ne 0)} 
 = \left. N_{\bm{r}} \right|_{(\{ p'_i \},a_k=0)}~.
\label{Nr2}
\end{eqnarray}

From (\ref{Nr1}) and (\ref{Nr2}),
we obtain the relation,
\begin{eqnarray}
  \left. N_{\bm{r}} \right|_{(\{ p_i \},a_k = 0)} 
 = \left. N_{\bm{r}} \right|_{(\{ p_i \},a_k \ne 0)}~,
\label{Nr3}
\end{eqnarray}
and find that each flavor number obtained from $[N,k]$ does not change   
even though the vacuum changes different ones in the presence of the Wilson line phases.

In this way, we arrive at the conjecture that
 {\it each flavor number in the SM
is independent of the Wilson line phases that respect the SM gauge group.}
If there were a Wilson line phase with a non-vanishing SM gauge quantum number,
(a part of) the SM gauge symmetry can be broken down.
Hence, we assume that such a Wilson line phase is vanishing or switched off.

\section{Fermion numbers and hidden supersymmetry}
\label{Fermion numbers and hidden supersymmetry}

On a higher-dimensional space-time $M^4 \times K^{D-4}$, 
a massless fermion $\Psi =\Psi(x,y)$ satisfies the equation,
\begin{eqnarray}
i\Gamma^M D_M \Psi = 0~,
\label{D-eq}
\end{eqnarray}
where $K^{D-4}$ is an $(D-4)$-dimensional extra space,
$\Gamma^M$ $(M=0, 1, 2, 3, 5, \cdots, D)$ are matrices that satisfy 
the Clifford algebra $\Gamma^M \Gamma^N + \Gamma^N \Gamma^M = 2 \eta^{MN}$,
$D_M \equiv \partial_M+ig A_M$ and
$\Psi$ is a fermion with $2^{[D/2]}$-components.
Here, $g$ is a gauge coupling constant, $A_M (=A_M^{\alpha} T^{\alpha})$ are gauge bosons,
and $[*]$ is the Gauss symbol.
The coordinates $x^{\mu}$ $(\mu = 0,1,2,3)$ on $M^4$ 
and $x^m$ $(m=5, \cdots, D)$ on $K^{D-4}$
are denoted by $x$ and $y$, respectively.

After the breakdown of gauge symmetry,
$\Psi$ is decomposed as
\begin{eqnarray}
\Psi(x, y) 
= \sum_{\bm{r}_H} \sum_{\{n_i\}} \left[\psi_{{\rm L}\bm{r}_H}^{\{n_i\}}(x) 
\phi_{{\rm L}\bm{r}_H}^{\{n_i\}}(y)
+ \psi_{{\rm R}\bm{r}_H}^{\{n_i\}}(x) \phi_{{\rm R}\bm{r}_H}^{\{n_i\}}(y)\right]~,
\label{Psi-decomp}
\end{eqnarray}
where $\psi_{{\rm L}\bm{r}_H}^{\{n_i\}}(x)$ and $\psi_{{\rm R}\bm{r}_H}^{\{n_i\}}(x)$
are 4D left-handed spinors and right-handed ones, respectively.
The subscript $\bm{r}_H$ stands for some representation of the unbroken gauge group $H$,
and the superscript $\{n_i\}$ represents a set of numbers relating massive modes 
and those concerning components of multiplet $\bm{r}_H$.
The functions $\phi_{{\rm L}\bm{r}_H}^{\{n_i\}}(y)$ and $\phi_{{\rm R}\bm{r}_H}^{\{n_i\}}(y)$
form complete sets on $K^{D-4}$.

We define the chiral fermion number relating $\bm{r}$ as
\begin{eqnarray}
n_{\bm{r}} \equiv n^0_{{\rm L}\bm{r}} - n^0_{{\rm R}\bm{r}}~,
\label{nr-def}
\end{eqnarray}
where $\bm{r}$ is a representation of the subgroup
unbroken in the presence of the Wilson line phases.
The net chiral fermion number $N_{\bm{r}}$
is given by $N_{\bm{r}} = n_{\bm{r}} - n_{\overline{\bm{r}}}$.

In case that $n_{\bm{r}}$ is independent of the Wilson line phases $(2\pi a_k)$,
$n^0_{{\rm L}\bm{r}}$ and $n^0_{{\rm R}\bm{r}}$ must be expressed as
\begin{eqnarray}
 n^0_{{\rm L}\bm{r}} =  n'^0_{{\rm L}\bm{r}} + f_{\bm{r}}(a_k)~~~{\rm and}~~~
 n^0_{{\rm R}\bm{r}} =  n'^0_{{\rm R}\bm{r}} + f_{\bm{r}}(a_k)~,
\label{nLR}
\end{eqnarray}
respectively.
Here, $n'^0_{{\rm L}\bm{r}}$ and  $n'^0_{{\rm R}\bm{r}}$ are some constants
irrelevant to $a_k$ and $f_{\bm{r}}(a_k)$ is a function of $a_k$.

\subsection{An example}
\label{An example}

Let us calculate $n^0_{{\rm L}\bm{r}}$ and $n^0_{{\rm R}\bm{r}}$,
and verify the relations (\ref{nLR}),
using an $SU(3)$ gauge theory on $M^4 \times S^1/Z_2$.

On 5D space-time, $\Psi$ is expressed as
\begin{eqnarray}
  \Psi = \left( \begin{array}{c} \psi_{\rm L} \\ \psi_{\rm R} \end{array} \right)~,
\label{psi}
\end{eqnarray}
where $\psi_{\rm L}$ and $\psi_{\rm R}$ are components 
containing 4D left-handed fermions and 4D right-handed ones, respectively.

The equation (\ref{D-eq}) is divided into two parts,
\begin{eqnarray}
  i\overline{\sigma}^{\mu}D_{\mu} \psi_{\rm L} - D_y\psi_{\rm R} = 0~,~~
  i\sigma^{\mu}D_{\mu} \psi_{\rm R} + D_y\psi_{\rm L} = 0~,
\label{5D-eqs}
\end{eqnarray}
where $D_y \equiv \partial_y + i g A_y$.
For $\psi_{\rm L}$ and $\psi_{\rm R}$, the BCs are given by
\begin{eqnarray}
&~&  \psi_{\rm L}(x,-y) = \eta^0 P_0 \psi_{\rm L}(x,y)~,~~
  \psi_{\rm L}(x,2\pi R-y) = \eta^1 P_1 \psi_{\rm L}(x,y)~,
\label{BC-L}\\
&~&  \psi_{\rm R}(x,-y) = -\eta^0 P_0 \psi_{\rm R}(x,y)~,~~
\psi_{\rm R}(x,2\pi R-y) = -\eta^1 P_1 \psi_{\rm R}(x,y)~,
 \label{BC-R}
 \end{eqnarray}
where $P_0$ and $P_1$ are the representation matrices for the $Z_2$ 
transformation $y \to -y$ 
and the $Z_2$ transformation $y \to 2\pi R -y$, respectively.
$\eta^0$ and $\eta^1$ are the intrinsic $Z_2$ parities for the left-handed component.
Note that $Z_2$ parities for the right-handed one are opposite to those of the left-handed one.
For the gauge bosons, the BCs are given by
\begin{eqnarray}
&~&  A_{\mu}(x,-y) = P_0 A_{\mu}(x,y)P_0^{\dagger}~,~~
A_{\mu}(x,2\pi R-y) = P_1 A_{\mu}(x,y)P_1^{\dagger}~,
\label{BC-Amu}\\
&~& A_y(x,-y) = -P_0 A_y(x,y)P_0^{\dagger}~,~~
A_y(x,2\pi R-y) = -P_1 A_y(x,y)P_1^{\dagger}~.
\label{BC-Ay}
 \end{eqnarray}

We take the representation matrices,
\begin{eqnarray}
  P_0 = \text{diag}(1,1,-1)~,~~
  P_1 = \text{diag}(1,1,-1)~.
\label{P++-}
\end{eqnarray}
Then $SU(3)$ is broken down to $SU(2) \times U(1)$.
We consider the fermion with the representation $\bf{3}$ of $SU(3)$ 
and $(\eta^0,\eta^1)=(1,1)$.
Then, $\psi_{\rm L}$ and $\psi_{\rm R}$ are expanded as
\begin{eqnarray}
  \psi_{\rm L} 
= \left( \begin{array}{c} \textstyle\sum\limits^{\infty}_{n=0} \psi^1_{{\rm L}n}(x)\cos{\frac{n}{R}y} \\ 
                                 \textstyle\sum\limits^{\infty}_{n=0} \psi^2_{{\rm L}n}(x)\cos{\frac{n}{R}y} \\ 
                                 \textstyle\sum\limits^{\infty}_{n=1} \psi^3_{{\rm L}n}(x)\sin{\frac{n}{R}y} 
\end{array} \right)~,~~
  \psi_{\rm R} 
= \left( \begin{array}{c} \textstyle\sum\limits^{\infty}_{n=1} \psi^1_{{\rm R}n}(x)\sin{\frac{n}{R}y} \\ 
                                 \textstyle\sum\limits^{\infty}_{n=1} \psi^2_{{\rm R}n}(x)\sin{\frac{n}{R}y} \\ 
                                 \textstyle\sum\limits^{\infty}_{n=0} \psi^3_{{\rm R}n}(x)\cos{\frac{n}{R}y} 
\end{array} \right). 
\label{Exp++-}
\end{eqnarray}

After a suitable $SU(2)$ gauge transformation, 
the vacuum expectation value (VEV) of $A_y$ is parameterized as
\begin{equation}
 \langle A_y \rangle 
= \frac{-i}{gR} \left( \begin{array}{ccc} 0&0&a \\ 0&0&0 \\ -a&0&0 \end{array} \right)~,
\label{Ay++-}
\end{equation}
where $2 \pi a$ is the Wilson line phase.
From the periodicity, we limit the domain of definition for $a$ as $0 \le a < 1$.
In case with $a \ne 0$, $SU(2)$ is broken down to $U(1)$,
and then every 4D fermion becomes a singlet.

Inserting (\ref{Exp++-}) and (\ref{Ay++-}) into (\ref{5D-eqs}),
we obtain a set of 4D equations,
\begin{eqnarray}
&~&   i\overline{\sigma}^{\mu}D_{\mu} \psi_{{\rm L}0}^1 
- \frac{a}{R} \psi_{{\rm R}0}^3 = 0~,~~ 
i\sigma^{\mu}D_{\mu} \psi_{{\rm R}0}^3
- \frac{a}{R} \psi_{{\rm L}0}^1 = 0~,
\label{DLR0}\\
&~&   i\overline{\sigma}^{\mu}D_{\mu} \psi_{{\rm L}0}^2 = 0~, 
\label{DLR0-2}\\
&~&  i\overline{\sigma}^{\mu}D_{\mu}\psi_{{\rm L}n}^1 
- \frac{n}{R}\psi_{{\rm R}n}^1 - \frac{a}{R}\psi_{{\rm R}n}^3 = 0 ~~~~(n=1,2,\cdots)~,
\label{DL1} \\
&~&  i\overline{\sigma}^{\mu}D_{\mu}\psi_{{\rm L}n}^2 - \frac{n}{R}\psi_{{\rm R}n}^2 = 0 
~~~~~~~~~~~~~~~~~~~~~(n=1,2,\cdots)~,
\label{DL2} \\
&~&  i\overline{\sigma}^{\mu}D_{\mu}\psi_{{\rm L}n}^3 + \frac{n}{R}\psi_{{\rm R}n}^3 
+ \frac{a}{R}\psi_{{\rm R}n}^1 = 0 
~~~~(n=1,2,\cdots)~,
\label{DL3} \\
&~&  i\sigma^{\mu}D_{\mu} \psi_{{\rm R}n}^1 - \frac{n}{R}\psi_{{\rm L}n}^1 
+ \frac{a}{R}\psi_{{\rm L}n}^3 = 0 
~~~~(n=1,2,\cdots)~,
\label{DR1} \\
&~&  i\sigma^{\mu}D_{\mu} \psi_{{\rm R}n}^2 - \frac{n}{R}\psi_{{\rm L}n}^2 =0 
~~~~~~~~~~~~~~~~~~~~~(n=1,2,\cdots)~,
\label{DR2} \\
&~&   i\sigma^{\mu}D_{\mu} \psi_{{\rm R}n}^3 + \frac{n}{R}\psi_{{\rm L}n}^3 
- \frac{a}{R}\psi_{{\rm L}n}^1 = 0 ~~~~(n=1,2,\cdots)~.
\label{DR3}
\end{eqnarray}
Using the equations (\ref{DL1}), (\ref{DL3}), (\ref{DR1}) and (\ref{DR3}),
we derive a set of 4D equations,
\begin{eqnarray}
&~&   i\overline{\sigma}^{\mu}D_{\mu}(\psi_{{\rm L}n}^1+\psi_{{\rm L}n}^3) 
- \frac{n-a}{R}(\psi_{{\rm R}n}^1-\psi_{{\rm R}n}^3) = 0 ~~~~(n=1,2,\cdots)~,
\label{DL1+3} \\
&~&   i\overline{\sigma}^{\mu}D_{\mu}(\psi_{{\rm L}n}^1-\psi_{{\rm L}n}^3) 
- \frac{n+a}{R}(\psi_{{\rm R}n}^1+\psi_{{\rm R}n}^3) = 0 ~~~~(n=1,2,\cdots)~,
\label{DL1-3} \\
&~&   i\sigma^{\mu}D_{\mu} (\psi_{{\rm R}n}^1+\psi_{{\rm R}n}^3) 
- \frac{n+a}{R}(\psi_{{\rm L}n}^1-\psi_{{\rm L}n}^3) = 0 ~~~~(n=1,2,\cdots)~,
\label{DR1+3} \\
&~&   i\sigma^{\mu}D_{\mu} (\psi_{{\rm R}n}^1-\psi_{{\rm R}n}^3) 
- \frac{n-a}{R}(\psi_{{\rm L}n}^1+\psi_{{\rm L}n}^3) = 0 ~~~~(n=1,2,\cdots)~.
\label{DR1-3}
\end{eqnarray}

From  (\ref{DLR0}), $\psi_{{\rm L}0}^1$ and $\psi_{{\rm R}0}^3$ form 
a 4D Dirac fermion.
In the same way, we find that
$(\psi_{{\rm L}n}^2, \psi_{{\rm R}n}^2)$, 
$(\psi_{{\rm L}n}^1 + \psi_{{\rm L}n}^3, \psi_{{\rm R}n}^1- \psi_{{\rm R}n}^3)$
and $(\psi_{{\rm L}n}^1 - \psi_{{\rm L}n}^3, \psi_{{\rm R}n}^1 + \psi_{{\rm R}n}^3)$ 
form 4D Dirac fermions for $n=1, 2, \cdots$
from (\ref{DL2}) and (\ref{DR2}), (\ref{DL1+3}) and (\ref{DR1-3}),
and (\ref{DL1-3}) and (\ref{DR1+3}), respectively.
.

\begin{figure}[h]
\begin{center}
\begin{picture}(350,140)
\linethickness{1pt}
\put(-10,-4){$0$}
\put(-23,26){$1/R$}
\put(-23,56){$2/R$}
\put(-23,86){$3/R$}
\put(27,110){\hbox{$\vdots$}}
\put(25,-20){$\psi_{\rm L}$}
\multiput(0,0)(0,30){4}{\line(1,0){60}}
\multiput(15,0)(0,30){4}{\circle*{5}}
\multiput(30,0)(0,30){4}{\circle*{5}}
\multiput(45,30)(0,30){3}{\circle*{5}}

\put(117,110){\hbox{$\vdots$}}
\put(115,-20){$\psi_{\rm R}$}
\multiput(90,0)(0,30){4}{\line(1,0){60}}
\multiput(105,0)(0,30){4}{\circle{5}}
\multiput(120,30)(0,30){3}{\circle{5}}
\multiput(135,30)(0,30){3}{\circle{5}}

\put(50,-45){$(a)~~a = 0$}

\put(190,-4){$0$}
\put(177,26){$1/R$}
\put(177,56){$2/R$}
\put(177,86){$3/R$}
\put(227,110){\hbox{$\vdots$}}
\put(225,-20){$\psi_{\rm L}$}
\multiput(200,0)(0,30){4}{\line(1,0){60}}
\multiput(200,0)(0,30){3}{\multiput(0,23)(6,0){10}{\line(1,0){4}}}
\multiput(200,0)(0,30){4}{\multiput(0,7)(6,0){10}{\line(1,0){4}}}
\multiput(215,0)(0,30){4}{\circle*{5}}
\multiput(245,23)(0,30){3}{\circle*{5}}
\multiput(230,7)(0,30){4}{\circle*{5}}

\put(350,4){\footnotesize $a/R$}
\put(350,20){\footnotesize $(1-a)/R$}
\put(350,34){\footnotesize $(1+a)/R$}
\put(350,50){\footnotesize $(2-a)/R$}
\put(350,64){\footnotesize $(2+a)/R$}
\put(350,80){\footnotesize $(3-a)/R$}
\put(350,94){\footnotesize $(3+a)/R$}
\put(317,110){\hbox{$\vdots$}}
\put(315,-20){$\psi_{\rm R}$}
\multiput(290,0)(0,30){4}{\line(1,0){60}}
\multiput(290,0)(0,30){3}{\multiput(0,23)(6,0){10}{\line(1,0){4}}}
\multiput(290,0)(0,30){4}{\multiput(0,7)(6,0){10}{\line(1,0){4}}}
\multiput(305,30)(0,30){3}{\circle{5}}
\multiput(335,23)(0,30){3}{\circle{5}}
\multiput(320,7)(0,30){4}{\circle{5}}

\put(250,-45){$(b)~~a \ne 0$}

\end{picture}
\end{center}
\abovecaptionskip=40pt
\caption{Mass spectrum of 4D fermions. 
The filled circles and the open ones represent left-handed fermions 
and right-handed ones, respectively.}
\label{Fig1}
\end{figure}
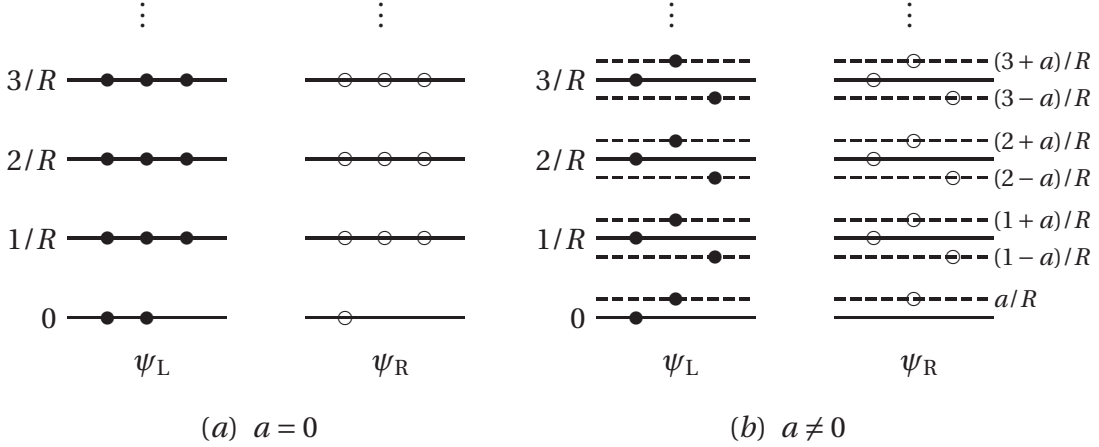

The numbers of 4D massless fermions are evaluated as
\begin{eqnarray}
  n^0_{\rm L} = 1 + \delta_{0 a}~,~~
  n^0_{\rm R}=  \delta_{0 a}~,
\label{nLR-ex}
\end{eqnarray}
where $\delta_{0 a}$ represents the Kronecker delta.
From (\ref{nLR-ex}), we confirm that 
the fermion number $n(\equiv n^0_{\rm L}-n^0_{\rm R}= 1)$ 
does not depend on the Wilson line phase.
The mass spectrum for 4D fermions in this model is depicted as Figure \ref{Fig1}.
Strictly speaking, the figure describes the case with $0 < a < 1/2$.

\subsection{Hidden quantum-mechanical supersymmetry}
\label{Hidden quantum-mechanical supersymmetry}

We explore a physics behind the feature
that the fermion numbers are independent of the Wilson line phases.

From Figure \ref{Fig1},
we anticipate that the feature originates from a hidden quantum-mechanical SUSY.
Here, the quantum-mechanical SUSY means the symmetry generated by the supercharge 
$Q$ that satisfies the algebraic relations~\cite{Witten1,CK&S},
\begin{eqnarray}
H = Q^2~,~~  \left\{ Q , (-1)^F \right\} = 0~,~~ \left((-1)^F\right)^2 = I~,
\label{SUSY}
\end{eqnarray}
where $H$, $F$ and $I$ are the Hamiltonian, the $\lq\lq$fermion'' number operator
and the identity operator, respectively.
The eigenvalue of $(-1)^F$ is given by $+1$ for $\lq\lq$bosonic'' states 
and $-1$ for $\lq\lq$fermionic'' states,
and $\text{Tr}~(-1)^F$ is a topological invariant, called the Witten index~\cite{Witten2}.

It is known that the system with 4D fermions has the hidden SUSY 
where the 4D Dirac operator plays the role of $Q$~\cite{Al,F&W}.
The correspondences are given by
\begin{eqnarray}
  Q \leftrightarrow i\gamma^{\mu}D_{\mu} = \left( \begin{array}{cc} 0 &  i{\sigma}^{\mu}D_{\mu} \\
   i\overline{\sigma}^{\mu}D_{\mu} & 0 \end{array} \right)~,~~
  (-1)^F \leftrightarrow \gamma_5~,
\label{SUSY-Dirac}
\end{eqnarray}
where $\gamma_5$ is the chirality operator defined by 
$\gamma_5 \equiv i \gamma^0 \gamma^1 \gamma^2 \gamma^3$.
The trace of $\gamma_5$ is the index of the 4D Dirac operator, and 
the following relations hold,
\begin{eqnarray}
&~& \left. \text{Tr}~\gamma_5\right|_{\bm{r}} 
 = n^0_{{\rm R}\bm{r}}[A_{\mu}] - n^0_{{\rm L}\bm{r}}[A_{\mu}]
 = \dim \ker {\sigma}^{\mu}D_{\mu}|_{\bm{r}} 
- \dim \ker \overline{\sigma}^{\mu}D_{\mu}|_{\bm{r}}
\nonumber \\
&~& ~~~~~~~~~~~~~~ = \frac{1}{32\pi^2}\int \text{tr}_{\bm{r}} \epsilon_{\mu\nu\alpha\beta}
F^{\mu\nu}F^{\alpha\beta}d^4x~,
\label{ASindex}
\end{eqnarray}
from the Atiyah-Singer index theorem.
Here, $n^0_{{\rm R}\bm{r}}[A_{\mu}]$ and $n^0_{{\rm L}\bm{r}}[A_{\mu}]$
are the numbers of 
normalizable solutions (massless fermions) 
satisfying $ i{\sigma}^{\mu}D_{\mu} \psi_{{\rm R}\bm{r}} = 0$
and $ i\overline{\sigma}^{\mu}D_{\mu} \psi_{{\rm L}\bm{r}} = 0$, respectively.
Note that massive fermions exist in pairs ($\psi_{{\rm R}\bm{r}}$ and $\psi_{{\rm L}\bm{r}}$)
and do not contribute to the index.
The integral quantity in (\ref{ASindex}) is called the Pontryagin number,
and it is deeply connected to the configuration of gauge bosons $A_{\mu}$ on 4D space-time.

It is pointed out that higher-dimensional theories with extra dimensions also
possess the hidden SUSY~\cite{LNS&S,S}.
In the system with a 5D fermion, the Dirac operator relating the fifth-coordinate 
plays the role of $Q$ and there are the correspondences,
\begin{eqnarray}
  Q \leftrightarrow \tilde{D}_y = \left( \begin{array}{cc} 0 & D_y \\ -D_y & 0 \end{array} \right)~,~~
  (-1)^F \leftrightarrow \tilde{\Gamma} \equiv \left( \begin{array}{cc} 1 & 0 \\ 0 & -1 \end{array} \right)~. 
\label{SUSY-Dy}
\end{eqnarray}
Note that $\tilde{\Gamma} = -\gamma_5$.
The counterpart of the Witten index is given by
\begin{eqnarray}
\left. \text{Tr}~\tilde{\Gamma}\right|_{\bm{r}} 
= \tilde{n}^0_{{\rm R}\bm{r}}(a) - \tilde{n}^0_{{\rm L}\bm{r}}(a)~, 
\label{TrGamma}
\end{eqnarray}
where $\tilde{n}^0_{{\rm R}\bm{r}}(a)$ and $\tilde{n}^0_{{\rm L}\bm{r}}(a)$ are 
the numbers of eigenfunctions, 
that satisfy the equations,
\begin{eqnarray}
  \tilde{D}_y \left( \begin{array}{c} 0 \\ \psi_{\rm R} \end{array} \right) 
= \left( \begin{array}{c} D_y\psi_{\rm R} \\ 0 \end{array} \right) 
= \left( \begin{array}{c} 0 \\ 0 \end{array} \right)
\label{DyR}
\end{eqnarray}
and
\begin{eqnarray}
  \tilde{D}_y \left( \begin{array}{c} \psi_{\rm L} \\ 0 \end{array} \right) 
= \left( \begin{array}{c} 0 \\ -D_y \psi_{\rm L} \end{array} \right) 
= \left( \begin{array}{c} 0 \\ 0 \end{array} \right)~,
\label{DyL}
\end{eqnarray}
respectively.
Note that the eigenvalue equations are given by $D_y \psi_{\rm R} = \lambda \psi_{\rm R}$
and $D_y \psi_{\rm L} = \lambda' \psi_{\rm L}$,
eigenfunctions with non-zero eigenvalues exist in pairs,
which correspond to 4D massive fermions as seen from (\ref{5D-eqs}),
and they do not contribute to the index.
From the equations (\ref{5D-eqs}),
there is a one-to-one correspondence such that
\begin{eqnarray}
  D_y \psi_{\rm R} = 0 \leftrightarrow i \overline{\sigma}^{\mu} D_{\mu} \psi_{\rm L} = 0~,~~
  D_y \psi_{\rm L} = 0 \leftrightarrow i \sigma^{\mu} D_{\mu} \psi_{\rm R} = 0~.
\label{Dy-SlashD}
\end{eqnarray}

Let us generalize to a system with a fermion on a higher-dimensional space-time.

For the case that $D=2n$ $(n=3, 4, \cdots)$, 
the correspondences are given by
\begin{eqnarray}
  Q \leftrightarrow \tilde{D} \equiv \sum_{m=5}^{D} i \Gamma^m D_m~,~~
  (-1)^F \leftrightarrow \tilde{\Gamma} \equiv -\Gamma_{D+1}~,
 \label{SUSY-Dm-even}
\end{eqnarray}
where $\Gamma_{D+1}$ is the chirality operator defined by
$\displaystyle{\Gamma_{D+1} = (-i)^{n+1} \Gamma^0 \Gamma^1 \cdots \Gamma^D}$.

For the case that $D=2n+1$ $(n=2, 3, \cdots)$, 
the correspondences are given by
\begin{eqnarray}
  Q \leftrightarrow \tilde{D} \equiv U^{\dagger} \sum_{m=5}^{D} i \Gamma^m D_m U~,~~
  (-1)^F \leftrightarrow \tilde{\Gamma} \equiv i \Gamma^{D}~,
 \label{SUSY-Dm-odd}
\end{eqnarray}
where $U$ is the unitary matrix that satisfies the relation
$i\Gamma^{D} = U^{\dagger} \Gamma^1 U$,
and $i \Gamma^{D}$ is a diagonal matrix with the same form as the chirality operator
on $D(=2n)$-dimensions up to a sign factor.

The equation (\ref{D-eq}) is written by
\begin{eqnarray}
i\Gamma^{\mu} D_{\mu} \Psi + \sum_{m=5}^{D} i\Gamma^m D_m \Psi = 0~.
\label{D-eqs}
\end{eqnarray}
For the case that $D=2n+1$, after the unitary transformation
$\Gamma'^M = U^{\dagger} \Gamma^M U$ and $\Psi' = U^{\dagger} \Psi$
is performed,  $\Gamma'^M$ and $\Psi'$ are again denoted as $\Gamma^M$ and $\Psi$
in (\ref{D-eqs}).
The counterpart of the Witten index is given by
\begin{eqnarray}
\left. \text{Tr}~\tilde{\Gamma}\right|_{\bm{r}} 
= \tilde{n}^0_{{\rm R}\bm{r}}(a_k) - \tilde{n}^0_{{\rm L}\bm{r}}(a_k)~, 
\label{TrGamma-ak}
\end{eqnarray}
where $\tilde{n}^0_{{\rm R}\bm{r}}(a_k)$ and $\tilde{n}^0_{{\rm L}\bm{r}}(a_k)$ 
are the numbers of eigenfunctions, 
that satisfy
$\tilde{D} \psi_{\rm R} = 0$ and $\tilde{D} \psi_{\rm L} = 0$, respectively.
From (\ref{D-eqs}), there is a one-to-one correspondence such that
\begin{eqnarray}
 \tilde{D} \psi_{\rm R} = 0 \leftrightarrow i \Gamma^{\mu} D_{\mu} \psi_{\rm L} = 0~,~~
 \tilde{D} \psi_{\rm L} = 0 \leftrightarrow i \Gamma^{\mu} D_{\mu} \psi_{\rm R} = 0~.
\label{tildeD-SlashD}
\end{eqnarray}
Here $\psi_{\rm R}$ and $\psi_{\rm L}$ are a 4D right-handed spinor component
and a 4D left-handed one in $\Psi$, 
that are eigenspinors of the 4D chirality operator 
$\Gamma_5 \equiv i \Gamma^0 \Gamma^1 \Gamma^2 \Gamma^3$
whose eigenvalues are $1$ and $-1$, respectively.
Note that components with a different 4D chirality involve each other 
through the equation (\ref{D-eqs}),
because $\Gamma_5$
is anti-commutable to $i\Gamma^{\mu} D_{\mu}$
but it is commutable to $\tilde{D}$.

From (\ref{tildeD-SlashD}),
the following relations hold,
\begin{eqnarray}
\tilde{n}^0_{{\rm R}\bm{r}}(a_k) = n^0_{{\rm L}\bm{r}}~,~~
\tilde{n}^0_{{\rm L}\bm{r}}(a_k) = n^0_{{\rm R}\bm{r}}~,
\label{tilde-n=n}
\end{eqnarray}
and, using (\ref{tilde-n=n}), we derive the relation,
\begin{eqnarray}
  \left. \text{Tr}~\tilde{\Gamma}\right|_{\bm{r}} 
= \tilde{n}^0_{{\rm R}\bm{r}}(a_k) - \tilde{n}^0_{{\rm L}\bm{r}}(a_k) 
= n^0_{{\rm L}\bm{r}} - n^0_{{\rm R}\bm{r}}~.
\label{TrGamma-2}
\end{eqnarray}
Because $\displaystyle{\left. \text{Tr}~\tilde{\Gamma}\right|_{\bm{r}}}$ is a topological invariant
and the Wilson line phases determine the vacuum with $\langle F_{mn} \rangle = 0$
globally in our orbifold family unification models,
$n_{\bm{r}} (= n^0_{{\rm L}\bm{r}} - n^0_{{\rm R}\bm{r}})$ 
is independent of the Wilson line phases.
Hence, $N_{\bm{r}} (= n_{\bm{r}} - n_{\overline{\bm{r}}})$ 
is also independent of the Wilson line phases.

Finally, we give a comment on $\displaystyle{\left. \text{Tr}~\tilde{\Gamma}\right|_{\bm{r}}}$.
As seen from the Atiyah-Singer index theorem relating the Dirac operator for extra-dimensions,
fermion numbers are deeply connected to the topological structure on $K^{D-4}$
including the configurations of $A_m$ on $K^{D-4}$.
From this point of view, the family number has been studied
in the Kaluza-Klein theory~\cite{Witten-KK}
and superstring theory~\cite{CHS&W}.

\section{Conclusions}
\label{Conclusions}

We have studied the relationship between the family number of chiral fermions
and the Wilson line phases, based on the orbifold family unification.
We have found that flavor numbers are independent 
of the Wilson line phases relating extra-dimensional components of gauge bosons,
as far as the SM gauge symmetry is respected.
This feature originates from a hidden quantum-mechanical SUSY.

From our observation, the previous analyses~\cite{KK&O,K&M,GK&M},
based on the assumption that the BCs are physical ones,
are justified in the orbifold family unification.
Concretely,
even if the BCs are not physical, we can obtain the same result
as that of the physical ones,
because the family number is invariant under the change
from the original BCs to the physical ones by singular gauge transformations.

\section*{Acknowledgments}
This work was supported in part by scientific grants from the Ministry of Education, Culture,
Sports, Science and Technology under Grant Nos.~22540272 and 21244036 (Y.K.).

\appendix

\section{Formulas based on equivalence relations}

We present several formulas concerning the combination ${}_nC_l$, 
derived from the dynamical rearrangement and
the feature that fermion numbers are independent of the Wilson line phases.

On $S^1/Z_2$, 
we consider the representation matrices given by
\begin{eqnarray}
&~& P_{0} 
= {\mathrm{diag}}([+1]_{p_1},[+1]_{p_2},[-1]_{p_3},[-1]_{p_4})~,
\label{P0} \\
&~& P_{1} 
= {\mathrm{diag}}([+1]_{p_1},[-1]_{p_2},[+1]_{p_3},[-1]_{p_4})~,
\label{P1}
\end{eqnarray}
where $[\pm 1]_{p_i}$ represents $\pm 1$ for all $p_i$ elements.
Then, the following breakdown of $SU(N)$ gauge symmetry occurs:
\begin{eqnarray}
SU(N) \to SU(p_1)\times SU(p_2) \times SU(p_3) \times SU(p_4) \times U(1)^{3-m}~.
\label{Z2-GSB}
\end{eqnarray}
The $Z_2$ parities or BCs specified by integers $\{p_i\}$ are also 
denoted $[p_1; p_2, p_3; p_4]$.

After the breakdown of $SU(N)$, $[N, k]$
is decomposed as~\cite{KK&O}
\begin{eqnarray}
[N, k] = \sum_{l_1 =0}^{k} \sum_{l_2 = 0}^{k-l_1} \sum_{l_3 = 0}^{k-l_1-l_2}  
\left({}_{p_1}C_{l_1}, {}_{p_2}C_{l_2}, {}_{p_3}C_{l_3}, {}_{p_4}C_{l_4}\right)~,
\label{Nk-Z2}
\end{eqnarray}
where $p_4=N-p_1-p_2-p_3$, $l_4=k-l_1- l_2 -l_3$,
and we use ${}_{p} C_{l}$ instead of $[p, l]$.
Our notation is that ${}_{p}C_{l} = 0$ for $l > p$ and $l < 0$.

The $Z_2$ parities of
$\left({}_{p_1}C_{l_1}, {}_{p_2}C_{l_2}, {}_{p_3}C_{l_3}, {}_{p_4}C_{l_4}\right)$ for 4D left-handed
fermions are given by
\begin{eqnarray}
&~& \mathcal{P}_0 = (-1)^{l_3 + l_4} \eta_k^0 =  (-1)^{l_1 + l_2} (-1)^k \eta_k^0
=  (-1)^{l_1 + l_2 +\alpha}~, \\
&~& \mathcal{P}_1 = (-1)^{l_2 + l_4} \eta_k^1 =  (-1)^{l_1 + l_3} (-1)^k \eta_k^1
=  (-1)^{l_1 + l_3 +\beta}~,
\label{mathcalP}
\end{eqnarray}
where the intrinsic $Z_2$ parities $(\eta_k^0, \eta_k^1)$ 
take a value $+1$ or $-1$ by definition
and are parameterized as $(-1)^{k}\eta^0_{k} = (-1)^{\alpha}$ and
$(-1)^{k}\eta^1_{k} = (-1)^{\beta}$.

Zero modes for the left-handed fermions 
and the right-handed ones are picked out by operating the projection
operators,
\begin{eqnarray}
P^{(1, 1)} = \frac{1+ \mathcal{P}_0}{2} \frac{1+ \mathcal{P}_1}{2}~~~~~
{\rm and}~~~~~
P^{(-1, -1)} = \frac{1- \mathcal{P}_0}{2} \frac{1- \mathcal{P}_1}{2}~,
\label{P11}
\end{eqnarray}
respectively.
Note that the intrinsic $Z_2$ parities for the right-handed fermions are opposite to
those for the left-handed ones.

Then, the fermion number is given by
\begin{eqnarray}
n = n^0_{\rm L} -n^0_{\rm R} = \sum_{l_1 =0}^{k} \sum_{l_2 = 0}^{k-l_1} \sum_{l_3 = 0}^{k-l_1-l_2} 
\left(P^{(1,1)} - P^{(-1,-1)}\right)
{}_{p_1}C_{l_1}~ {}_{p_2}C_{l_2}~ {}_{p_3}C_{l_3}~ {}_{p_4}C_{l_4}~.
\label{n-S1/Z2}
\end{eqnarray}

From the dynamical rearrangement, the following equivalence relations hold~\cite{HH&K},
\begin{eqnarray}
&~& [p_1; p_2, p_3; p_4] 
 \sim [p_1-1; p_2+1, p_3+1; p_4-1]~~~~ (\mbox{for}~~ p_1, p_4 \ge 1) ~,
\nonumber \\
&~& ~~~~~~~~~~~~~~~~~~~~~~~~~~~~ \sim [p_1+1; p_2-1, p_3-1; p_4+1]~~~~ (\mbox{for}~~ p_2, p_3 \ge 1) ~.
\label{equ-S1/Z2}
\end{eqnarray}
Using (\ref{equ-S1/Z2}) and 
the feature that fermion numbers are independent of the Wilson line phases,
the following formula is derived,
\begin{eqnarray}
&~& \sum_{l_1 =0}^{k} \sum_{l_2 = 0}^{k-l_1} \sum_{l_3 = 0}^{k-l_1-l_2} 
\left[(-1)^{l_1 + l_2 +\alpha} +  (-1)^{l_1 + l_3 +\beta}\right]
{}_{p_1}C_{l_1}~ {}_{p_2}C_{l_2}~ {}_{p_3}C_{l_3}~ {}_{p_4}C_{l_4}
\nonumber \\
&~& = \sum_{l_1 =0}^{k} \sum_{l_2 = 0}^{k-l_1} \sum_{l_3 = 0}^{k-l_1-l_2} 
\left[(-1)^{l_1 + l_2 +\alpha} +  (-1)^{l_1 + l_3 +\beta}\right]
{}_{p_1 \mp 1}C_{l_1}~ {}_{p_2 \pm 1}C_{l_2}~ {}_{p_3 \pm 1}C_{l_3}~ {}_{p_4 \mp 1}C_{l_4}~,
\label{F-S1/Z2}
\end{eqnarray}
where $p_4=N-p_1-p_2-p_3$, $l_4=k-l_1- l_2 -l_3$,
and we use the relation,
\begin{eqnarray}
P^{(1, 1)} -  P^{(-1,-1)} 
= \frac{1}{2}\left(\mathcal{P}_0 + \mathcal{P}_1\right)
= \frac{1}{2}\left[ (-1)^{l_1 + l_2 +\alpha}+  (-1)^{l_1 + l_3 +\beta}\right]~.
\label{P11-P-1-1}
\end{eqnarray}
Here and hereafter, we deal with the case 
that the inequality $p_i - 1 \ge 0$ is fulfilled in ${}_{p_i-1} C_{l_i}$.

In the same way, the following formulas are derived 
from the feature of the fermion number on $T^2/Z_2$,
\begin{eqnarray}
&~& \sum_{l_1 =0}^{k} \sum_{l_2 = 0}^{k-l_1} \sum_{l_3 =0}^{k-l_1-l_2} \sum_{l_4 = 0}^{k-l_1-l_2-l_3}
\sum_{l_5 = 0}^{k-l_1-\cdots -l_4}\sum_{l_6 = 0}^{k-l_1-\cdots -l_5} 
 \sum_{l_7 = 0}^{k-l_1-\cdots -l_6} 
\left(P^{(1,1,1)} - P^{(-1,-1,-1)}\right)
\nonumber \\
&~& ~~~~~~~~~~~~~~~~~~~~~~~~~~~~~~~~~~ 
\times {}_{p_1}C_{l_1}~ {}_{p_2}C_{l_2}~ {}_{p_3}C_{l_3}~ {}_{p_4}C_{l_4}~
{}_{p_5}C_{l_5}~ {}_{p_6}C_{l_6}~ {}_{p_7}C_{l_7}~ {}_{p_8}C_{l_8}
\nonumber \\
&~& = \sum_{l_1 =0}^{k} \sum_{l_2 = 0}^{k-l_1} \sum_{l_3 =0}^{k-l_1-l_2} \sum_{l_4 = 0}^{k-l_1-l_2-l_3}
\sum_{l_5 = 0}^{k-l_1-\cdots -l_4}\sum_{l_6 = 0}^{k-l_1-\cdots -l_5} 
 \sum_{l_7 = 0}^{k-l_1-\cdots -l_6} 
\left(P^{(1,1,1)} - P^{(-1,-1,-1)}\right)
\nonumber \\
&~& ~~~~~~~~~~~~~~~~~~~~~~~~~~~~~~~~~~ 
\times {}_{p_1 \mp 1}C_{l_1}~ {}_{p_2 \pm 1}C_{l_2}~ {}_{p_3}C_{l_3}~ {}_{p_4}C_{l_4}~
{}_{p_5}C_{l_5}~ {}_{p_6}C_{l_6}~ {}_{p_7 \pm 1}C_{l_7}~ {}_{p_8 \mp 1}C_{l_8}
\nonumber \\
&~& = \sum_{l_1 =0}^{k} \sum_{l_2 = 0}^{k-l_1} \sum_{l_3 =0}^{k-l_1-l_2} \sum_{l_4 = 0}^{k-l_1-l_2-l_3}
\sum_{l_5 = 0}^{k-l_1-\cdots -l_4}\sum_{l_6 = 0}^{k-l_1-\cdots -l_5} 
 \sum_{l_7 = 0}^{k-l_1-\cdots -l_6} 
\left(P^{(1,1,1)} - P^{(-1,-1,-1)}\right)
\nonumber \\
&~& ~~~~~~~~~~~~~~~~~~~~~~~~~~~~~~~~~~ 
\times {}_{p_1}C_{l_1}~ {}_{p_2 \mp 1}C_{l_2}~ {}_{p_3 \pm 1}C_{l_3}~ {}_{p_4}C_{l_4}~
{}_{p_5}C_{l_5}~ {}_{p_6 \pm 1}C_{l_6}~ {}_{p_7 \mp 1}C_{l_7}~ {}_{p_8}C_{l_8}
\nonumber \\
&~& = \sum_{l_1 =0}^{k} \sum_{l_2 = 0}^{k-l_1} \sum_{l_3 =0}^{k-l_1-l_2} \sum_{l_4 = 0}^{k-l_1-l_2-l_3}
\sum_{l_5 = 0}^{k-l_1-\cdots -l_4}\sum_{l_6 = 0}^{k-l_1-\cdots -l_5} 
 \sum_{l_7 = 0}^{k-l_1-\cdots -l_6} 
\left(P^{(1,1,1)} - P^{(-1,-1,-1)}\right)
\nonumber \\
&~& ~~~~~~~~~~~~~~~~~~~~~~~~~~~~~~~~~~ 
\times {}_{p_1}C_{l_1}~ {}_{p_2 \mp 1}C_{l_2}~ {}_{p_3}C_{l_3}~ {}_{p_4 \pm 1}C_{l_4}~
{}_{p_5 \pm 1}C_{l_5}~ {}_{p_6}C_{l_6}~ {}_{p_7 \mp 1}C_{l_7}~ {}_{p_8}C_{l_8}
\nonumber \\
&~& = \sum_{l_1 =0}^{k} \sum_{l_2 = 0}^{k-l_1} \sum_{l_3 =0}^{k-l_1-l_2} \sum_{l_4 = 0}^{k-l_1-l_2-l_3}
\sum_{l_5 = 0}^{k-l_1-\cdots -l_4}\sum_{l_6 = 0}^{k-l_1-\cdots -l_5} 
 \sum_{l_7 = 0}^{k-l_1-\cdots -l_6} 
\left(P^{(1,1,1)} - P^{(-1,-1,-1)}\right)
\nonumber \\
&~& ~~~~~~~~~~~~~~~~~~~~~~~~~~~~~~~~~~ 
\times {}_{p_1}C_{l_1}~ {}_{p_2}C_{l_2}~ {}_{p_3 \pm 1}C_{l_3}~ {}_{p_4 \mp 1}C_{l_4}~
{}_{p_5 \mp 1}C_{l_5}~ {}_{p_6 \pm 1}C_{l_6}~ {}_{p_7}C_{l_7}~ {}_{p_8}C_{l_8}~,
\label{F-T2/Z2}
\end{eqnarray}
where $p_8=N-p_1-p_2-\cdots - p_7$ and $l_8=k-l_1- l_2 -\cdots - l_7$.
$P^{(a, b, c)}$ are the projection operators
that pick out the $Z_2$ parities $(\mathcal{P}_0, \mathcal{P}_1, \mathcal{P}_2)=(a, b, c)$,
defined by
\begin{eqnarray}
P^{(a, b, c)} \equiv \frac{1+ a \mathcal{P}_0}{2}\frac{1+ b \mathcal{P}_1}{2}
\frac{1+ c \mathcal{P}_2}{2}~.
\label{Pabc}
\end{eqnarray}
Here, $a$, $b$ and $c$ take $1$ or $-1$.
$\mathcal{P}_0$,  $\mathcal{P}_1$ and $\mathcal{P}_2$ are given by
\begin{eqnarray}
&~& \mathcal{P}_{0} 
= (-1)^{l_5+l_6+l_7+l_8} \eta^0_{k} = (-1)^{l_1+l_2+l_3+l_4} (-1)^k \eta^0_{k} 
= (-1)^{l_1+l_2+l_3+l_4+\alpha}~, 
\label{Z20}\\
&~& \mathcal{P}_{1} 
= (-1)^{l_3+l_4+l_7+l_8} \eta^1_{k} = (-1)^{l_1+l_2+l_5+l_6} (-1)^k \eta^1_{k}
= (-1)^{l_1+l_2+l_5+l_6+\beta}~,
\label{Z21}\\
&~& \mathcal{P}_{2} 
= (-1)^{l_2+l_4+l_6+l_8} \eta^2_{k} = (-1)^{l_1+l_3+l_5+l_7} (-1)^k \eta^2_{k}
= (-1)^{l_1+l_3+l_5+l_7+\gamma}~,
\label{Z22}
\end{eqnarray}
where $\alpha$, $\beta$ and $\gamma$ take $0$ or $1$.
Using (\ref{Z20}), (\ref{Z21}) and (\ref{Z22}), $P^{(1,1,1)} - P^{(-1,-1,-1)}$
is calculated as
\begin{eqnarray}
&~& P^{(1, 1, 1)} -  P^{(-1,-1,-1)} 
\nonumber \\
&~& ~~~ = \frac{1}{4}\left[(-1)^{l_1+l_2+l_3+l_4 +\alpha} + (-1)^{l_1+l_2+l_5+l_6 + \beta} \right.
\nonumber \\
&~& ~~~~~~~~~~~~~ 
\left. + (-1)^{l_1+ l_3+l_5+l_7+\gamma} 
+ (-1)^{l_1+l_4+l_6+l_7 +\alpha+\beta+\gamma} \right]~.
\label{P111-P-1-1-1}
\end{eqnarray}

The following formulas are derived
from the feature of the fermion numbers relating representations ${}_{p_1} C_{l_1}$
and $({}_{p_1} C_{l_1}, {}_{p_2} C_{l_2})$,
\begin{eqnarray}
&~& \sum_{l_2 = 0}^{k-l_1} \sum_{l_3 =0}^{k-l_1-l_2} \sum_{l_4 = 0}^{k-l_1-l_2-l_3}
\sum_{l_5 = 0}^{k-l_1-\cdots -l_4}\sum_{l_6 = 0}^{k-l_1-\cdots -l_5} 
 \sum_{l_7 = 0}^{k-l_1-\cdots -l_6} 
\left(P^{(1,1,1)} - P^{(-1,-1,-1)}\right)
\nonumber \\
&~& ~~~~~~~~~~~~~~~~~~~~~~~~~~~~~~~~~~~~~~ 
\times {}_{p_2}C_{l_2}~ {}_{p_3}C_{l_3}~ {}_{p_4}C_{l_4}~
{}_{p_5}C_{l_5}~ {}_{p_6}C_{l_6}~ {}_{p_7}C_{l_7}~ {}_{p_8}C_{l_8}
\nonumber \\
&~& = \sum_{l_2 = 0}^{k-l_1} \sum_{l_3 =0}^{k-l_1-l_2} \sum_{l_4 = 0}^{k-l_1-l_2-l_3}
\sum_{l_5 = 0}^{k-l_1-\cdots -l_4}\sum_{l_6 = 0}^{k-l_1-\cdots -l_5} 
 \sum_{l_7 = 0}^{k-l_1-\cdots -l_6} 
\left(P^{(1,1,1)} - P^{(-1,-1,-1)}\right)
\nonumber \\
&~& ~~~~~~~~~~~~~~~~~~~~~~~~~~~~~~~~~~~~~~ 
\times {}_{p_2 \mp 1}C_{l_2}~ {}_{p_3 \pm 1}C_{l_3}~ {}_{p_4}C_{l_4}~
{}_{p_5}C_{l_5}~ {}_{p_6 \pm 1}C_{l_6}~ {}_{p_7 \mp 1}C_{l_7}~ {}_{p_8}C_{l_8}
\nonumber \\
&~& = \sum_{l_2 = 0}^{k-l_1} \sum_{l_3 =0}^{k-l_1-l_2} \sum_{l_4 = 0}^{k-l_1-l_2-l_3}
\sum_{l_5 = 0}^{k-l_1-\cdots -l_4}\sum_{l_6 = 0}^{k-l_1-\cdots -l_5} 
 \sum_{l_7 = 0}^{k-l_1-\cdots -l_6} 
\left(P^{(1,1,1)} - P^{(-1,-1,-1)}\right)
\nonumber \\
&~& ~~~~~~~~~~~~~~~~~~~~~~~~~~~~~~~~~~~~~~ 
\times {}_{p_2 \mp 1}C_{l_2}~ {}_{p_3}C_{l_3}~ {}_{p_4 \pm 1}C_{l_4}~
{}_{p_5 \pm 1}C_{l_5}~ {}_{p_6}C_{l_6}~ {}_{p_7 \mp 1}C_{l_7}~ {}_{p_8}C_{l_8}
\nonumber \\
&~& = \sum_{l_2 = 0}^{k-l_1} \sum_{l_3 =0}^{k-l_1-l_2} \sum_{l_4 = 0}^{k-l_1-l_2-l_3}
\sum_{l_5 = 0}^{k-l_1-\cdots -l_4}\sum_{l_6 = 0}^{k-l_1-\cdots -l_5} 
 \sum_{l_7 = 0}^{k-l_1-\cdots -l_6} 
\left(P^{(1,1,1)} - P^{(-1,-1,-1)}\right)
\nonumber \\
&~& ~~~~~~~~~~~~~~~~~~~~~~~~~~~~~~~~~~~~~~ 
\times {}_{p_2}C_{l_2}~ {}_{p_3 \pm 1}C_{l_3}~ {}_{p_4 \mp 1}C_{l_4}~
{}_{p_5 \mp 1}C_{l_5}~ {}_{p_6 \pm 1}C_{l_6}~ {}_{p_7}C_{l_7}~ {}_{p_8}C_{l_8}
\label{F2-T2/Z2}
\end{eqnarray}
and 
\begin{eqnarray}
&~& \sum_{l_3 =0}^{k-l_1-l_2} \sum_{l_4 = 0}^{k-l_1-l_2-l_3}
\sum_{l_5 = 0}^{k-l_1-\cdots -l_4}\sum_{l_6 = 0}^{k-l_1-\cdots -l_5} 
 \sum_{l_7 = 0}^{k-l_1-\cdots -l_6} 
\left(P^{(1,1,1)} - P^{(-1,-1,-1)}\right)
\nonumber \\
&~& ~~~~~~~~~~~~~~~~~~~~~~~~~~~~~~~~~~~~~~ 
\times {}_{p_3}C_{l_3}~ {}_{p_4}C_{l_4}~
{}_{p_5}C_{l_5}~ {}_{p_6}C_{l_6}~ {}_{p_7}C_{l_7}~ {}_{p_8}C_{l_8}
\nonumber \\
&~& =\sum_{l_3 =0}^{k-l_1-l_2} \sum_{l_4 = 0}^{k-l_1-l_2-l_3}
\sum_{l_5 = 0}^{k-l_1-\cdots -l_4}\sum_{l_6 = 0}^{k-l_1-\cdots -l_5} 
 \sum_{l_7 = 0}^{k-l_1-\cdots -l_6} 
\left(P^{(1,1,1)} - P^{(-1,-1,-1)}\right)
\nonumber \\
&~& ~~~~~~~~~~~~~~~~~~~~~~~~~~~~~~~~~~~~~~ 
\times {}_{p_3 \pm 1}C_{l_3}~ {}_{p_4 \mp 1}C_{l_4}~
{}_{p_5 \mp 1}C_{l_5}~ {}_{p_6 \pm 1}C_{l_6}~ {}_{p_7}C_{l_7}~ {}_{p_8}C_{l_8}~.
\label{F3-T2/Z2}
\end{eqnarray}

Furthermore,
by changing $(p_3, p_4, p_5, p_6, p_7, p_8)$ into $(p_7, p_8, p_3, p_4, p_5, p_6)$
in the ordering of the summation
and relabeling $(p_7, p_8, p_3, p_4, p_5, p_6)$ as $(p_3, p_4, p_5, p_6, p_7, p_8)$, 
the following formulas are derived from the feature of the fermion numbers relating 
representations $({}_{p_1} C_{l_1}, {}_{p_2} C_{l_2}, {}_{p_3} C_{l_3})$
and $({}_{p_1} C_{l_1}, {}_{p_2} C_{l_2}, {}_{p_3} C_{l_3}, {}_{p_4} C_{l_4})$,
\begin{eqnarray}
&~& \sum_{l_4 = 0}^{k-l_1-l_2-l_3}
\sum_{l_5 = 0}^{k-l_1-\cdots -l_4}\sum_{l_6 = 0}^{k-l_1-\cdots -l_5} 
 \sum_{l_7 = 0}^{k-l_1-\cdots -l_6} 
\left(P'^{(1,1,1)} - P'^{(-1,-1,-1)}\right)
\nonumber \\
&~& ~~~~~~~~~~~~~~~~~~~~~~~~~~~~~~~~~~~~~~ 
\times {}_{p_4}C_{l_4}~
{}_{p_5}C_{l_5}~ {}_{p_6}C_{l_6}~ {}_{p_7}C_{l_7}~ {}_{p_8}C_{l_8}
\nonumber \\
&~& = \sum_{l_4 = 0}^{k-l_1-l_2-l_3}
\sum_{l_5 = 0}^{k-l_1-\cdots -l_4}\sum_{l_6 = 0}^{k-l_1-\cdots -l_5} 
 \sum_{l_7 = 0}^{k-l_1-\cdots -l_6} 
\left(P'^{(1,1,1)} - P'^{(-1,-1,-1)}\right)
\nonumber \\
&~& ~~~~~~~~~~~~~~~~~~~~~~~~~~~~~~~~~~~~~~ 
\times {}_{p_4}C_{l_4}~
{}_{p_5 \mp 1}C_{l_5}~ {}_{p_6 \pm 1}C_{l_6}~ {}_{p_7 \pm 1}C_{l_7}~ {}_{p_8 \mp 1}C_{l_8}
\label{F4-T2/Z2}
\end{eqnarray}
and
\begin{eqnarray}
\hspace{-1.3cm}&~& \sum_{l_5 = 0}^{k-l_1-\cdots -l_4}\sum_{l_6 = 0}^{k-l_1-\cdots -l_5} 
 \sum_{l_7 = 0}^{k-l_1-\cdots -l_6} 
\left(P'^{(1,1,1)} - P'^{(-1,-1,-1)}\right)
{}_{p_5}C_{l_5}~ {}_{p_6}C_{l_6}~ {}_{p_7}C_{l_7}~ {}_{p_8}C_{l_8}
\nonumber \\
\hspace{-1.3cm}&~& = \sum_{l_5 = 0}^{k-l_1-\cdots -l_4}\sum_{l_6 = 0}^{k-l_1-\cdots -l_5} 
 \sum_{l_7 = 0}^{k-l_1-\cdots -l_6} 
\left(P'^{(1,1,1)} - P'^{(-1,-1,-1)}\right)
{}_{p_5 \mp 1}C_{l_5}~ {}_{p_6 \pm 1}C_{l_6}~ {}_{p_7 \pm 1}C_{l_7}~ {}_{p_8 \mp 1}C_{l_8}~,
\label{F5-T2/Z2}
\end{eqnarray}
where $P'^{(1, 1, 1)} -  P'^{(-1,-1,-1)}$ is given by
\begin{eqnarray}
&~& P'^{(1, 1, 1)} -  P'^{(-1,-1,-1)} 
\nonumber \\
&~& ~~~ = \frac{1}{4}\left[(-1)^{l_1+l_2+l_5+l_6 +\alpha} + (-1)^{l_1+l_2+l_7+l_8 + \beta} \right.
\nonumber \\
&~& ~~~~~~~~~~~~~ \left. + (-1)^{l_1+ l_3+l_5+l_7+\gamma} 
+ (-1)^{l_1+l_3+l_6+l_8 +\alpha+\beta+\gamma} \right]~.
\label{P111-P-1-1-1-prime}
\end{eqnarray}

In the same way, the following formulas are derived 
from the feature of the fermion number on $T^2/Z_3$,
\begin{eqnarray}
&~& \sum_{l_1 =0}^{k} \sum_{l_2 = 0}^{k-l_1} \sum_{l_3 =0}^{k-l_1-l_2} \sum_{l_4 = 0}^{k-l_1-l_2-l_3}
\sum_{l_5 = 0}^{k-l_1-\cdots -l_4}\sum_{l_6 = 0}^{k-l_1-\cdots -l_5} 
 \sum_{l_7 = 0}^{k-l_1-\cdots -l_6} \sum_{l_8 = 0}^{k-l_1-\cdots -l_7} 
\left(P^{(1,1)} - P^{(\omega, \omega)}\right)
\nonumber \\
&~& ~~~~~~~~~~~~~~~~~~~~~ 
\times {}_{p_1}C_{l_1}~ {}_{p_2}C_{l_2}~ {}_{p_3}C_{l_3}~ {}_{p_4}C_{l_4}~
{}_{p_5}C_{l_5}~ {}_{p_6}C_{l_6}~ {}_{p_7}C_{l_7}~ {}_{p_8}C_{l_8}~ {}_{p_9}C_{l_9}
\nonumber \\
&~& = \sum_{l_1 =0}^{k} \sum_{l_2 = 0}^{k-l_1} \sum_{l_3 =0}^{k-l_1-l_2} \sum_{l_4 = 0}^{k-l_1-l_2-l_3}
\sum_{l_5 = 0}^{k-l_1-\cdots -l_4}\sum_{l_6 = 0}^{k-l_1-\cdots -l_5} 
 \sum_{l_7 = 0}^{k-l_1-\cdots -l_6}  \sum_{l_8 = 0}^{k-l_1-\cdots -l_7}
\left(P^{(1,1)} - P^{(\omega, \omega)}\right)
\nonumber \\
&~& ~~~~~~~~~~~~~~~~~~~~~ 
\times {}_{p_1 \pm 1}C_{l_1}~ {}_{p_2}C_{l_2}~ {}_{p_3 \mp 1}C_{l_3}~ {}_{p_4 \mp 1}C_{l_4}~
{}_{p_5 \pm 1}C_{l_5}~ {}_{p_6}C_{l_6}~ {}_{p_7}C_{l_7}~ {}_{p_8 \mp 1}C_{l_8}~ {}_{p_9 \pm 1}C_{l_9}
\nonumber \\
&~& = \sum_{l_1 =0}^{k} \sum_{l_2 = 0}^{k-l_1} \sum_{l_3 =0}^{k-l_1-l_2} \sum_{l_4 = 0}^{k-l_1-l_2-l_3}
\sum_{l_5 = 0}^{k-l_1-\cdots -l_4}\sum_{l_6 = 0}^{k-l_1-\cdots -l_5} 
 \sum_{l_7 = 0}^{k-l_1-\cdots -l_6}  \sum_{l_8 = 0}^{k-l_1-\cdots -l_7}
\left(P^{(1,1)} - P^{(\omega, \omega)}\right)
\nonumber \\
&~& ~~~~~~~~~~~~~~~~~~~~~ 
\times {}_{p_1 \pm 1}C_{l_1}~ {}_{p_2 \mp 1}C_{l_2}~ {}_{p_3}C_{l_3}~ {}_{p_4}C_{l_4}~
{}_{p_5 \pm 1}C_{l_5}~ {}_{p_6 \mp 1}C_{l_6}~ {}_{p_7 \mp 1}C_{l_7}~ {}_{p_8}C_{l_8}~ {}_{p_9 \pm 1}C_{l_9}
\nonumber \\
&~& = \sum_{l_1 =0}^{k} \sum_{l_2 = 0}^{k-l_1} \sum_{l_3 =0}^{k-l_1-l_2} \sum_{l_4 = 0}^{k-l_1-l_2-l_3}
\sum_{l_5 = 0}^{k-l_1-\cdots -l_4}\sum_{l_6 = 0}^{k-l_1-\cdots -l_5} 
 \sum_{l_7 = 0}^{k-l_1-\cdots -l_6}  \sum_{l_8 = 0}^{k-l_1-\cdots -l_7}
\left(P^{(1,1)} - P^{(\omega, \omega)}\right)
\nonumber \\
&~& ~~~~~~~~~~~~~~~~~~~~~ 
\times {}_{p_1}C_{l_1}~ {}_{p_2 \pm 1}C_{l_2}~ {}_{p_3 \mp 1}C_{l_3}~ {}_{p_4 \mp 1}C_{l_4}~
{}_{p_5}C_{l_5}~ {}_{p_6 \pm 1}C_{l_6}~ {}_{p_7 \pm 1}C_{l_7}~ {}_{p_8 \mp 1}C_{l_8}~ {}_{p_9}C_{l_9}~,
\label{F-T2/Z3}
\end{eqnarray}
where $p_9=N-p_1-p_2-\cdots - p_8$ and $l_9 =k-l_1- l_2 -\cdots - l_8$.
$P^{(\xi, \eta)}$ are the projection operators
that pick out the $Z_3$ elements $({\Theta}_0, {\Theta}_1)=(\xi, \eta)$,
defined by
\begin{eqnarray}
P^{(\xi, \eta)} \equiv \frac{1+ \overline{\xi} {\Theta}_0 + \overline{\xi}^2 {\Theta}_0^2}{3} 
\frac{1+ \overline{\eta} {\Theta}_1 + \overline{\eta}^2 {\Theta}_1^2}{3}~.
\label{Pxy}
\end{eqnarray}
Here, $\xi$ and $\eta$ take $1$, $\omega(=e^{2\pi i/3})$ or $\overline{\omega}(= e^{4\pi i/3})$,
and $\overline{\xi}$ and $\overline{\eta}$ 
are the complex conjugates of $\xi$ and $\eta$, respectively.
${\Theta}_0$ and  ${\Theta}_1$ are given by
\begin{eqnarray}
&~& {\Theta}_{0} = \omega^{l_4+l_5+l_6} \overline{\omega}^{l_7+l_8+l_9} \eta^0_{k}
= \omega^{l_1+l_2+l_3+ 2(l_4+l_5+l_6)} \overline{\omega}^{k} \eta^0_{k}
= \omega^{l_1+l_2+l_3+ 2(l_4+l_5+l_6) + \alpha} ~, 
\label{Z30}\\
&~& {\Theta}_{1} = \omega^{l_2+l_5+l_8} \overline{\omega}^{l_3+l_6+l_9} \eta^1_{k}
= \omega^{l_1+l_4+l_7+ 2(l_2+l_5+l_8)} \overline{\omega}^{k} \eta^1_{k}
= \omega^{l_1+l_4+l_7+ 2(l_2+l_5+l_8) + \beta}~,
\label{Z31}
\end{eqnarray}
where $\alpha$ and $\beta$ take $0$, $1$ or $2$.

In the same way, we can derive similar formulas from the feature of the fermion numbers relating 
representations ${}_{p_1} C_{l_1}$, $({}_{p_1} C_{l_1}, {}_{p_2} C_{l_2})$
and $({}_{p_1} C_{l_1}, {}_{p_2} C_{l_2}, {}_{p_3} C_{l_3})$
on $T^2/Z_3$.

\section{Formulas based on independence from Wilson line phases}

We derive other formulas concerning the combination ${}_nC_l$,
counting the numbers of fermions irrelevant to the Wilson line phases
and
using the independence of fermion numbers from the Wilson line phases.

On $S^1/Z_2$, 
we consider the representation matrices given by
\begin{eqnarray}
P_{0} 
= {\mathrm{diag}}([+1]_{p},[-1]_{N-p})~,~~
P_{1} 
= {\mathrm{diag}}([+1]_{p},[-1]_{N-p})~.
\label{P1-ps}
\end{eqnarray}
Then, the following breakdown of $SU(N)$ gauge symmetry occurs:
\begin{eqnarray}
SU(N) \to SU(p)\times SU(N-p) \times U(1)^{1-m}~,
\label{Z2-GSB-ps}
\end{eqnarray}
and $[N, k]$ is decomposed as
\begin{eqnarray}
[N, k] = \sum_{l =0}^{k} 
\left({}_{p}C_{l}, {}_{N-p}C_{k-l}\right)~.
\label{Nk-Z2-ps}
\end{eqnarray}

The $Z_2$ parities of $\left({}_{p}C_{l}, {}_{s}C_{k-l}\right)$ for 4D left-handed
fermions are given and parameterized by
\begin{eqnarray}
\mathcal{P}_0 = (-1)^{k-l} \eta_k^0 
=  (-1)^{l +\alpha}~,~~
\mathcal{P}_1 = (-1)^{k-l} \eta_k^1 
=  (-1)^{l +\beta}~,
\label{mathcalP-ps}
\end{eqnarray}
where $\alpha$ and $\beta$ take $0$ or $1$.
Then, the fermion number is given by
\begin{eqnarray}
n = n_{\rm L} -n_{\rm R} = \sum_{l =0}^{k} \frac{1}{2}
\left[(-1)^{l +\alpha} +  (-1)^{l +\beta}\right]
 {}_{p}C_{l}~ {}_{N-p}C_{k-l}~.
\label{n-S1/Z2-ps}
\end{eqnarray}

The number of the Wilson line phases is $m \equiv {\rm Min}(p, N-p)$
and, after a suitable $SU(p) \times SU(N-p)$ gauge transformation,
$\langle A_y \rangle$ is parameterized as
\begin{equation}
 \langle A_y \rangle 
= \frac{-i}{gR} \left( \begin{array}{cc} 0&\Theta \\  -\Theta^{T}&0 \end{array} \right)~,
\label{<Ay>}
\end{equation}
where $\Theta$ is the $p \times (N- p)$ matrix such that
\begin{eqnarray}
&~& \Theta 
= \begin{pmatrix}
    \begin{matrix}
      a_1 &  \\
      & a_2 \\
      & \vphantom{\begin{matrix} \ddots \\ a_m  \end{matrix}} \raisebox{-5pt}{\LARGE{0}} \\
      0 & 0 \\
      \vdots & \vdots \\
      0 & 0 
    \end{matrix} &
    \begin{matrix}
      \vphantom{\begin{matrix} a_1 \\ a_2 \end{matrix}} \mbox{\LARGE{0}} & \\
      \ddots & \\
      & a_m \\
      \cdots & 0 \\
      \ddots & \vdots \\
      \cdots & 0
    \end{matrix}
\end{pmatrix}
~~~~~({\rm for}~~p \ge N-p)~,
\label{Theta-p}\\
&~& \Theta 
= \begin{pmatrix}
    \begin{matrix}
      a_1 & \\
      & a_2 \\
      & \vphantom{\begin{matrix} \vdots \\ 0 \end{matrix}} \raisebox{-5pt}{\LARGE{0}} \\
    \end{matrix} &
    \begin{matrix}
      \mbox{\LARGE{0}} & \vphantom{\begin{matrix} \cdots \\ 0 \end{matrix}} \\
      \ddots & \\
      & a_m \\
    \end{matrix} &
    \begin{matrix}
     \vphantom{a_1}0 & \cdots & 0 \\
     \vphantom{a_2}0 & \cdots & 0 \\
     \vdots & \ddots & \vdots \\
     \vphantom{a_m}0 & \cdots & 0 \\
    \end{matrix}
\end{pmatrix}
~~~~~({\rm for}~~p \le N-p)~.
\label{Theta-N-p}
\end{eqnarray}
Here, $2\pi a_k$ $(k=1, \cdots, m; m \equiv {\rm Min}(p, N-p))$ are the Wilson line phases.

For the fermion with $[N, 1]$, the number of components irrelevant to $a_k$
is $p-m$ for $p \ge N-p$ and $N-p -m$ for $p \le N-p$, and it is expressed as
\begin{equation}
\left. \sum_{l'=0}^1 {}_{p-m} C_{l'}~ {}_{N-p-m} C_{1-l'}\right|_{m = {\rm Min}(p, N-p)}~.
\label{[N,1]}
\end{equation}
For the fermion with $[N, 2]$, the number of components irrelevant to  $a_k$
is ${}_{p-m} C_2 + m$ for $p \ge N-p$ and ${}_{N-p-m} C_{2} +m$ for $p \le N-p$, 
and it is expressed as
\begin{equation}
\left. \sum_{l'=0}^2 {}_{p-m} C_{l'}~ {}_{N-p-m} C_{2-l'} +  {}_m C_{1}\right|_{m = {\rm Min}(p, N-p)}~,
\label{[N,2]}
\end{equation}
where ${}_m C_{1}$ comes from the components constructed from the tensor products
between components in $[N,1]$ with opposite values for the Wilson line phases,
and the components corresponding ${}_m C_{1}$ have odd $Z_2$ parities.
In the iterative fashion, we find that
the number of components irrelevant to $a_k$
is given by
\begin{eqnarray}
\left. \sum_{n=0}^{[k/2]} \sum_{l'=0}^{k-2n} {}_m C_n~ {}_{p-m}C_{l'}~ 
{}_{N-p-m}C_{k-2n-l'}\right|_{m = {\rm Min}(p, N-p)}
\label{S1/Z2-irr-Wilson}
\end{eqnarray}
for the fermion with $[N, k]$.

Using the independence of fermion numbers from the Wilson line phases,
the number of fermions is also calculated by counting the fermions
irrelevant to $a_k$ and the following formula is derived,
\begin{eqnarray}
 \sum_{l =0}^{k} (-1)^l {}_{p}C_{l}~ {}_{N-p}C_{k-l}
= \sum_{n=0}^{[k/2]} \sum_{l'=0}^{k-2n}  (-1)^{n+l'} {}_m C_n~ 
{}_{p-m}C_{l'}~ {}_{N-p-m}C_{k-2n-l'}~,
\label{S1/Z2-F1}
\end{eqnarray}
where we use the assignment of $Z_2$ parities,
\begin{eqnarray}
\mathcal{P}_0 = (-1)^{n+k-2n-l'} \eta_k^0 
=  (-1)^{n+l' +\alpha}~,~~
\mathcal{P}_1 = (-1)^{n+k-2n-l'} \eta_k^1 
=  (-1)^{n+l' +\beta}
\label{mathcalP-ps-l'}
\end{eqnarray}
for the component corresponding
${}_m C_n~ {}_{p-m}C_{l'}~ {}_{N-p-m}C_{k-2n-l'}$, and we take $\alpha = \beta$.
The above formula (\ref{S1/Z2-F1}) holds for the integer $m$ 
satisfying $0 \le m \le {\rm Min}(p, N-p)$,
because the above argument is valid for $m$ as the number of
non-vanishing $a_k$ even if some of $a_k$ vanish.

Particularly, in case with $m=p$ and $m=N-p$,  (\ref{S1/Z2-F1}) reduces to 
\begin{eqnarray}
\sum_{l =0}^{k} (-1)^l {}_{p}C_{l}~ {}_{N-p}C_{k-l}
= \sum_{n=0}^{[k/2]} \sum_{l'=0}^{k-2n}  (-1)^{n+l'} {}_p C_n~ {}_{N-2p}C_{k-2n-l'}~,
\label{S1/Z2-F1-m=p}
\end{eqnarray}
and 
\begin{eqnarray}
\sum_{l =0}^{k} (-1)^l {}_{p}C_{l}~ {}_{N-p}C_{k-l}
= \sum_{n=0}^{[k/2]} \sum_{l'=0}^{k-2n}  (-1)^{n+l'} {}_{N-p} C_n~ {}_{2p-N}C_{k-2n-l'}~,
\label{S1/Z2-F1-m=N-p}
\end{eqnarray}
respectively.

Based on the representation matrices (\ref{P0}) and (\ref{P1}),
the following formula is derived,
\begin{eqnarray}
&~& \sum_{l_1 =0}^{k} \sum_{l_2 = 0}^{k-l_1} \sum_{l_3 = 0}^{k-l_1-l_2} 
\left[(-1)^{l_1 + l_2 +\alpha} +  (-1)^{l_1 + l_3 +\beta}\right]
{}_{p_1}C_{l_1}~ {}_{p_2}C_{l_2}~ {}_{p_3}C_{l_3}~ {}_{p_4}C_{l_4}
\nonumber \\
&~& = \sum_{n=0}^{[k/2]} \sum_{n_1=0}^{n}
\sum_{l'_1 =0}^{k-2n} \sum_{l'_2 = 0}^{k-2n- l'_1} \sum_{l'_3 = 0}^{k-2n-l'_1-l'_2} 
\left[(-1)^{n+l'_1 + l'_2 +\alpha} +  (-1)^{n+ l'_1 + l'_3 +\beta}\right]
\nonumber \\
&~& ~~~~~~~~ \times {}_{m_1} C_{n_1}~ {}_{m_2} C_{n-n_1}~
{}_{p_1 - m_1}C_{l'_1}~ {}_{p_2-m_2}C_{l'_2}~ {}_{p_3 - m_2}C_{l'_3}~ 
{}_{p_4 - m_1}C_{l'_4}~,
\label{S1/Z2-F2}
\end{eqnarray}
where $p_4 = N-p_1-p_2-p_3$ and $l'_4 = k-2n-l'_1-l'_2-l'_3$.
The above formula (\ref{S1/Z2-F2}) holds for the integers $m_1$ and $m_2$
satisfying $0 \le m_1 \le {\rm Min}(p_1, p_4)$ and $0 \le m_2 \le {\rm Min}(p_2, p_3)$.

In the same way, we can derive similar formulas using models on $T^2/Z_M$.

\end{document}